\documentclass{article}
\usepackage[dvips]{graphicx}

\begin{document}
\title{Proposal of a Simple Method to Estimate Neutrino Oscillation \\
Probability and CP Violation in Matter}
\author{
\sc 
{Akira Takamura$^{1,2}$}\thanks{E-mail
address:takamura@eken.phys.nagoya-u.ac.jp} , 
{Keiichi Kimura$^1$}\thanks
{E-mail address:kimukei@eken.phys.nagoya-u.ac.jp} 
{and Hidekazu Yokomakura$^1$}\thanks{E-mail address:yoko@eken.phys.nagoya-u.ac.jp} 
\\
\\
{\small \it $^1$Department of Physics, Nagoya University,}
{\small \it Nagoya, 464-8602, Japan}\\
{\small \it $^2$Department of Mathematics, 
Toyota National College of Technology}\\
{\small \it Eisei-cho 2-1, Toyota-shi, 471-8525, Japan}}
\date{}
\maketitle
\begin{abstract}
We study neutrino oscillation within the framework of three generations 
in matter.
We propose a simple method to approximate 
the coefficients $A, B$ and $C$ 
which do not depend on the CP phase $\delta$ in the oscillation probability 
$P(\nu_e \to \nu_{\mu})=A\cos \delta + B\sin \delta +C$.
An advantage of our method is that an approximate formula of 
the coefficients $A,B$ and $C$ in arbitrary matter 
{\it without the usual first order perturbative calculations} 
of the small parameter $\Delta m_{21}^2/\Delta m_{31}^2$ or 
$\sin \theta_{13}$ can be derived.
Furthermore we show that all the approximate formulas for low,
intermediate and high energy regions given by other authors 
in constant matter can be easily derived from our formula. 
It means that our formula is applicable over a wide energy region.
\end{abstract}

\section{Introduction}

\hspace*{\parindent}
Recent experiments clarified that the solar neutrino deficit 
and the atmospheric neutrino anomaly are strong evidences 
for the neutrino oscillations with three generations.
The solar neutrino deficit is explained by
$\nu_e \to \nu_\mu$ oscillation \cite{solar} and 
the atmospheric neutrino anomaly is explained 
by $\nu_{\mu} \to \nu_{\tau}$ oscillation \cite{atm}. 
In the recent SNO \cite{SNO} and KamLAND experiments
\cite{KamLAND}, the solar neutrino problem has been solved by 
the large mixing angle (LMA) MSW solution \cite{MSW}. 
Furthermore the upper bound of $\theta_{13}$ is given by 
the CHOOZ experiment \cite{CHOOZ}. 
Thus, there are two small parameters 
\begin{eqnarray} 
\alpha \equiv \Delta m_{21}^2/\Delta m_{31}^2 \sim 0.03, 
\qquad  \sin \theta_{13} \leq 0.16.
\end{eqnarray}
The remaining problems are the determination of sign $\Delta m_{31}^2$,
the measurement of the 1-3 mixing angle $\theta_{13}$ and 
the CP phase $\delta$ \cite{Degeneracy}.
In the limit of vanishing mixing angle 
$\theta_{13}$ or vanishing mass squared difference $\Delta m^2_{21}$,
the CP violating effects in the oscillation probability disappear.
Therefore the magnitude of the two small parameters 
$\alpha$ 
and $\sin \theta_{13}$ controls the magnitude of the CP violation.
The LMA MSW solution in the solar neutrino problem 
has opened the possibility of the observation of CP violation 
in the lepton sector.
For this purpose, many long baseline neutrino experiments 
are planned \cite{lbl}.
\par
The matter effect received from the earth is important
in the long baseline neutrino experiments, because 
fake CP violation is induced due to matter effect.
The Preliminary Reference Earth Model (PREM) is well known 
as the model of the earth density and is usually used in analysis 
of long baseline experiments.
However, it has recently been pointed out in geophysical analysis 
of the matter density profile from J-PARC to Beijing \cite{Modeling} 
that the deviation from the PREM is rather large.
In this paper, we derive an approximate formula 
of neutrino oscillation probability 
without assuming any specific earth density models. 
\par
In constant matter, various approximate formulas have been proposed 
in low energy \cite{Koike2000, Yasuda1999, Minakata2000},
in intermediate energy \cite{AKS,AS,Sato2000} 
and in high energy regions \cite{Freund2000, Cervera, Freund}.
In the case that the matter density is not constant,
approximate formulas have been also derived 
in \cite{Akhmedov,Akhmedov0402,KS98K2K,OS,Miura0106,Miura0102,Brahmachari0303} 
using perturbative calculations to analyze the terrestrial matter effect.
However, the question of how to separate the genuine CP violation 
due to the leptonic CP phase from the fake CP violation induced 
by matter effect has not been investigated sufficiently 
in arbitrary matter. 
\par
The next step is to analyze the CP violating effects in more detail
in the case of non-constant matter density.
In order to obtain a hint for this problem, we will briefly review 
the approach applied in the solar neutrino problem.
It is difficult to derive the exact solutions for solar neutrino 
problem in three generations except for some special matter profile.
As an approach to derive the neutrino oscillation probability, 
a low energy approximate formula was proposed in \cite{Kuo1988}.
By averaging $\Delta m_{31}^2$, they derived the formula
\begin{eqnarray}
P^{(3)}(\nu_e \to \nu_e) 
= \cos^4 \theta_{13}P^{(2)}(\nu_e \to \nu_e) + \sin^4 \theta_{13}.
\label{Kuo-Pantareone}
\end{eqnarray}
This is a formula to reduce the calculation of 
the survival probability $P^{(3)}(\nu_e \to\nu_e)$ 
in three generations to that of $P^{(2)}(\nu_e \to \nu_e)$ 
in two generations.
Therefore this formula is called the reduction formula
\cite{Reduction-application}.
This reduction formula is useful for the analysis of 
solar neutrino experiments,
but it is not directly applicable to long baseline neutrino experiments 
planned in the future, because we cannot average $\Delta m_{31}^2$ 
in long baseline experiments.
Therefore we need to derive the reduction formula 
which is valid without averaging $\Delta m_{31}^2$.
\par
In a series of previous papers we have calculated the oscillation 
probability $P(\nu_e \to \nu_\mu)$.
In the papers \cite{Kimura} we have shown 
that the CP phase $\delta$ dependence 
of $P(\nu_e \to \nu_\mu)$ in constant matter is given in the form
\begin{eqnarray}
P(\nu_e \to \nu_\mu) = A \cos \delta + B \sin \delta + C
\label{CP-dependence}
\end{eqnarray}
and have derived an exact but simple expression 
for the coefficients $A, B$ and $C$.
In the next paper \cite{Yokomakura0207} we have presented 
a simple and general formula
which does not depend on the matter profile.
As a result we have concluded that the equation (\ref{CP-dependence}) 
is valid even in arbitrary matter.
However, in the case of non-constant matter density, 
there exist no closed-form expressions for the 
coefficients $A, B$ and $C$.
\par
In this paper we propose a simple method to derive the approximate 
formula of the coefficients $A, B$ and $C$ 
taking account of the small parameters $\alpha$ and 
$\sin \theta_{13}$.
The coefficients $A$ and $B$ are linear in $\alpha$ and 
$\sin\theta_{13}$.
These coefficients represent the genuine three flavor effect.
Therefore it has been considered that the first order perturbative 
calculations of $\alpha$ or $\theta_{13}$ are needed for the derivation 
of $A$ and $B$.
However it is possible to calculate $A, B$ and $C$ 
{\it without the usual first order perturbative calculations} 
of small parameter $\alpha$ or $\sin \theta_{13}$ in our method.
As we shall see later in section 2,
the reduction formula in arbitrary matter is derived as 
\begin{eqnarray}
A &\simeq& 2 {\rm Re}[{S}_{\mu e}^{\ell *} S^h_{\tau e}] c_{23} s_{23}, 
\label{reduction-1} \\
B &\simeq& - 2 {\rm Im}[{S}_{\mu e}^{\ell *} S^h_{\tau e}] c_{23} s_{23}, 
\label{reduction-2} \\
C &\simeq& |S^\ell_{\mu e}|^2 c_{23}^2 + |S^h_{\tau e}|^2 s_{23}^2,
\label{reduction-3}
\end{eqnarray}
where $S^\ell_{\mu e}$ and $S^h_{\tau e}$ are the oscillation amplitudes 
calculated in the following Hamiltonian, respectively
\begin{eqnarray}
H^\ell &=& O_{12}{\rm diag}(0,\Delta_{21},\Delta_{31}) O_{12}^T 
+ {\rm diag}(a(t),0,0), \\
H^h &=& O_{13}{\rm diag}(0,0,\Delta_{31})O_{13}^T 
+ {\rm diag}(a(t),0,0).
\end{eqnarray}
Here $\Delta_{ij}$ is defined by $\Delta_{ij}=\Delta m_{ij}^2/2E$,
$O_{ij}$ is the rotational matrix in the $ij$ plane and 
$a(t)$ is the matter potential.
Since both $H^\ell$ and $H^h$ are Hamiltonians in two generations, 
the equations (\ref{reduction-1})-(\ref{reduction-3}) 
are formulas in order to reduce the calculation of the 
coefficients $A, B$ and $C$ in three generations to the 
oscillation amplitudes in two generations.
Furthermore we show that other approximate formulas 
known in constant matter can be easily derived from our formula.
This means that our formula is applicable to a wide energy region.

\section{New Idea for an Approximate Formula}

\hspace*{\parindent}
In this section we propose a new idea to derive 
an approximate formula for neutrino oscillation probability.
At first we review a general framework for the oscillation 
probability in arbitrary matter. 
Next we introduce how to derive an approximate formula from this framework.
We also discuss the difference between our method and usual methods.  

\subsection{Review of General Formulation}

\hspace*{\parindent}
In this subsection we briefly review that the CP dependence of 
the oscillation probability $P(\nu_e \to \nu_\mu)$ is given 
in the form as 
$P(\nu_e \to \nu_\mu) = A \cos \delta + B \sin \delta + C$
in arbitrary matter.
More detailed calculation has been given 
in the papers \cite{Yokomakura0207}.
\par
The Hamiltonian in matter is given by
\begin{eqnarray}
H =  U {\rm diag}(0,\Delta_{21},\Delta_{31}) U^\dagger 
+ {\rm diag}(a(t),0,0),
\end{eqnarray}
where $U$ is the Maki-Nakagawa-Sakata (MNS) matrix \cite{MNS}.
Using the standard parametrization 
\begin{eqnarray}
U = O_{23} \Gamma O_{13} \Gamma^\dagger O_{12},
\end{eqnarray}
the Hamiltonian is written as 
\begin{eqnarray}
H = O_{23} \Gamma O_{13}O_{12}{\rm diag}
(0,\Delta_{21},\Delta_{31}) O_{12}^TO_{13}^T \Gamma^\dagger O_{23}^T
+ {\rm diag}(a(t),0,0),
\end{eqnarray}
where $\Gamma = {\rm diag}(1,1,e^{i\delta})$ is the phase matrix, 
It should be pointed out that the matter potential of 
the Hamiltonian contains only the $e$-$e$ component.
The Hamiltonian can be written  
in the form decomposing the CP phase $\delta$ 
and the 2-3 mixing angle $\theta_{23}$ as
\begin{eqnarray}
H = O_{23} \Gamma H' \Gamma^\dagger O_{23}^T, \label{Decomposition}
\end{eqnarray}
where $H'$ is defined as
\begin{eqnarray}
 H' 
= O_{13}O_{12}{\rm diag}(0,\Delta_{21},\Delta_{31}) O_{12}^TO_{13}^T 
+ {\rm diag}(a(t),0,0). \label{relation-Hamiltonian}
\end{eqnarray}
The amplitudes $S$ and $S'$ are given by substituting
 the relations (\ref{Decomposition}) 
and (\ref{relation-Hamiltonian}) into the following equations
\begin{eqnarray}
S = {\rm T}\exp\left\{-i\int^L_0 H(t)dt \right\}, 
\quad 
S' = {\rm T}\exp\left\{-i\int^L_0 H'(t)dt \right\}.
\end{eqnarray}
Then we obtain the amplitude for $\nu_\beta' \to \nu_\alpha'$
as the $\alpha$-$\beta$ component   
\begin{eqnarray}
S_{\alpha\beta} = (O_{23} \Gamma S' \Gamma^\dagger O_{23}^T)_{\alpha\beta}.
\end{eqnarray}
In particular when we choose $\mu$ and $e$ as $\alpha$ and $\beta$, 
the amplitude $S_{\mu e}$ is given by
\begin{eqnarray}
S_{\mu e} = S'_{\mu e}c_{23} + S'_{\tau e} s_{23}e^{i\delta}.
\end{eqnarray}
From this relation, the probability is calculated as 
\begin{eqnarray}
P(\nu_e \to \nu_\mu) &=& A \cos \delta + B \sin \delta + C, \\
A &=& 2 {\rm Re}[{S}_{\mu e}^{'*} S'_{\tau e}] c_{23} s_{23}, 
\label{Coefficient-A} \\
B &=& - 2 {\rm Im}[{S}_{\mu e}^{'*} S'_{\tau e}] c_{23} s_{23}, 
\label{Coefficient-B} \\
C &=& |S'_{\mu e}|^2 c_{23}^2 + |S'_{\tau e}|^2 s_{23}^2,
\label{Coefficient-C}
\end{eqnarray}
which is the exact formula in arbitrary matter derived 
in the previous paper \cite{Yokomakura0207}.

\subsection{Order Counting  of  $A, B$ and $C$ 
on $\alpha$ and $\sin \theta_{13}$}

\hspace*{\parindent}
In this subsection we study how the coefficients $A, B$ and $C$ 
defined in (\ref{Coefficient-A})-(\ref{Coefficient-C})
depend on $\alpha$ and $\sin \theta_{13}$.
Instead of $A, B$ and $C$, we study the dependence of 
$S'_{\mu e}$ and $S'_{\tau e}$ on $\alpha$ and $\sin \theta_{13}$
by taking the limit either $\theta_{13} \to 0$ or $\alpha \to 0$.
\par
At first, taking the limit $\theta_{13} \to 0$, the Hamiltonian 
reduces to  
\begin{eqnarray}
 H^\ell &=& \lim_{\theta_{13} \to 0} H' \\
&=& O_{12}{\rm diag}(0,\Delta_{21},\Delta_{31}) O_{12}^T 
+ {\rm diag}(a(t),0,0) \\
&=&
  \left(
 \begin{array}{ccc}
  \Delta_{21} s_{12}^2 + a(t)
  & \Delta_{21}s_{12}c_{12} 
  & 0 \\
  \Delta_{21}s_{12}c_{12} 
  & \Delta_{21} c_{12}^2 
  & 0 \\
  0
  & 0
  &  \Delta_{31} 
 \end{array}
 \right). \label{low-Hamiltonian}
\end{eqnarray}
This Hamiltonian expresses the fact that the third generation 
is separated from the first and the second generations.
We simply obtain the amplitude 
\begin{equation}
 S'_{\tau e} = 0
\end{equation}
from the Hamiltonian (\ref{low-Hamiltonian}).
It means that the order of $S'_{\tau e}$ is given by 
\begin{equation}
 S'_{\tau e} = O(\sin \theta_{13}). \label{order-S^h}
\end{equation}
for the case $\theta_{13} \neq 0$.
In the same way, taking the limit $\Delta_{21} \to 0$,
the Hamiltonian reduces to 
\begin{eqnarray}
 H^h &=& \lim_{\Delta_{21} \to 0} H' \\
&=& O_{13}{\rm diag}(0,0,\Delta_{31})O_{13}^T 
+ {\rm diag}(a(t),0,0) \\
&=&
  \left(
 \begin{array}{ccc}
   \Delta_{31} s_{13}^2 + a(t)
  & 0
  & \Delta_{31}s_{13}c_{13} \\
  0
  & 0
  & 0 \\
   \Delta_{31}s_{13}c_{13} 
  & 0
  &  \Delta_{31} c_{13}^2
 \end{array}
 \right). \label{high-Hamiltonian}
\end{eqnarray}
This Hamiltonian expresses the fact that the second generation 
is separated from the first and the third generations.
We simply obtain the amplitude
\begin{equation}
 S'_{\mu e} = 0
\end{equation}
from the Hamiltonian (\ref{high-Hamiltonian}).
It means that the order of $S'_{\mu e}$ is given by
\begin{equation}
 S'_{\mu e} = O(\alpha) \label{order-S^l}
\end{equation}
for the case $\alpha \neq 0$.
Finally, we conclude that the dependence of the coefficients 
$A, B$ and $C$ on $\alpha$ and $\sin \theta_{13}$ is given by
\begin{eqnarray}
A &=& 2 {\rm Re}[{S}_{\mu e}^{'*} S'_{\tau e}] c_{23} s_{23}
   = O(\alpha \sin \theta_{13}), \label{counting-A} \\
B &=& - 2 {\rm Im}[{S}_{\mu e}^{'*} S'_{\tau e}] c_{23} s_{23}
   = O(\alpha \sin \theta_{13}), \label{counting-B} \\
C &=& |S'_{\mu e}|^2 c_{23}^2 + |S'_{\tau e}|^2 s_{23}^2
   = O(\alpha^2) + O(\sin^2 \theta_{13}). \label{counting-C} 
\end{eqnarray}
Since both $A$ and $B$ vanish in the two flavor limit, 
either $\alpha \to 0$ or $\sin \theta_{13} \to 0$, 
this fact represents the genuine three flavor effect.
The coefficients $A$ and $B$ are doubly suppressed 
by these small parameters $\alpha$ and $\sin \theta_{13}$.
In Refs. \cite{Cervera,Freund}, this is pointed out
for the case of constant matter density. 
However these results (\ref{counting-A})-(\ref{counting-C}) are correct 
even in arbitrary matter profile.

\subsection{Main Result}

\hspace*{\parindent}
In this subsection, we propose a simple method to approximately calculate 
the amplitudes $S'_{\mu e}$ and $S'_{\tau e}$.
From the result of the previous subsection, 
the dependence of $S'_{\mu e}$ and $S'_{\tau e}$ 
on $\alpha$ and $\sin \theta_{13}$ is given by
\begin{eqnarray}
S'_{\mu e} = O(\alpha), \quad S'_{\tau e} = O(\sin \theta_{13}).
\label{order}
\end{eqnarray}
We expand both $S'_{\mu e}$ and $S'_{\tau e}$ in terms of two small parameters 
$\alpha$ and $\sin \theta_{13}$ as
\begin{eqnarray}
S'_{\mu e} 
&=& \Big(
 O(\alpha) + O(\alpha^2) + O(\alpha^3) + \cdots
 \Big)
+ \Big(
 O(\alpha\sin \theta_{13}) +  O(\alpha^2\sin \theta_{13}) 
+ \cdots
 \Big) \\
&=&  S^\ell_{\mu e}
+ O(\alpha\sin \theta_{13}) +  O(\alpha^2\sin \theta_{13})
+ \cdots, 
\label{expansion-1} \\
S'_{\tau e} 
&=& \Big(
 O(\sin \theta_{13}) + O(\sin^2 \theta_{13}) + \cdots
 \Big)
+ \Big(
 O(\alpha\sin \theta_{13}) +  O(\alpha^2\sin \theta_{13}) + \cdots
 \Big) \\
&=& S^h_{\tau e}
+ O(\alpha\sin \theta_{13}) + O(\alpha^2\sin \theta_{13}) 
+ \cdots, 
\label{expansion-2}
\end{eqnarray} 
where $S^\ell_{\mu e}$ and $S^h_{\tau e}$ are defined by
\begin{eqnarray}
S^\ell_{\mu e} &=& \lim_{\theta_{13} \to 0} S'_{\mu e},  \\
S^h_{\tau e} &=& \lim_{\alpha \to 0} S'_{\tau e}.
\end{eqnarray}
From (\ref{expansion-1}) and (\ref{expansion-2}) 
we can approximate the amplitudes as
\begin{eqnarray}
S'_{\mu e} &\simeq& S^\ell_{\mu e} \label{leading-1}, 
\label{low} \\
S'_{\tau e} &\simeq& S^h_{\tau e} \label{leading-2}.
\label{high}
\end{eqnarray}
The accuracy of this approximation is determined by 
the magnitude of the higher order terms 
on $\sin \theta_{13}$ and $\alpha$.
At present, the upper bound of $\sin \theta_{13}$ is given by 
the CHOOZ experiment.
In future experiments, when the value of $\theta_{13}$ will become 
smaller, the accuracy of the approximate formula can be better.
It is noted that the simple method introduced in this subsection 
does not depend on whether the matter density is constant or not.
We obtain the oscillation probability from the reduced amplitudes as
\begin{eqnarray}
P(\nu_e \to \nu_\mu) &=& A \cos \delta + B \sin \delta + C,
\label{approximate-P} \\
A &\simeq& 2 {\rm Re}[{S}_{\mu e}^{\ell *} S^h_{\tau e}] c_{23} s_{23},
\label{approximate-A} \\
B &\simeq& - 2 {\rm Im}[{S}_{\mu e}^{\ell *} S^h_{\tau e}] c_{23} s_{23},
\label{approximate-B} \\
C &\simeq& |S^\ell_{\mu e}|^2 c_{23}^2 + |S^h_{\tau e}|^2 s_{23}^2.
\label{approximate-C}
\end{eqnarray}
This formula is one of the main results obtained in this paper.
The advantage of this formula is as follows.
First, this formula is derived by using only two small 
parameters $\alpha$ and $\sin \theta_{13}$ without assuming 
a specific matter density model. 
Therefore, this formula is applicable to the case of the PREM, ak135f 
and so on.
Second, the reduction formula (\ref{Kuo-Pantareone})
applied to the solar neutrino problem is valid only
in the case that the averaging for $\Delta m^2_{31}$ 
is possible.
However, our formula is effective even in the case that 
the oscillation probability can not be averaged.
Namely, it is applicable to long baseline experiments.
Third, we derive this formula {\it without the first order 
perturbative calculations} of small parameter $\alpha$ or 
$\sin \theta_{13}$. This is the reason why our derivation is 
easier than usual perturbative methods proposed by other authors.
More detailed discussion is given in the next subsection.

\subsection{Comparison with Usual Perturbative Calculations}

\hspace*{\parindent}
In this subsection, we compare our method with usual 
perturbative methods and describe the advantage of our method clearly.
From the result of the previous subsection, the dependence 
of the coefficients $A$ and $B$ on $\alpha$ and 
$\sin\theta_{13}$ is given by
\begin{eqnarray}
A = O(\alpha \sin \theta_{13}), \qquad 
B = O(\alpha \sin \theta_{13}).
\end{eqnarray}
As both $\alpha$ and $\sin\theta_{13}$ are small parameters, 
there are two kinds of perturbative methods. 
One method is to consider $\alpha$ as a small parameter 
and treat $\theta_{13}$ exactly.
Another method is to consider $\sin\theta_{13}$ as 
a small parameter and treat $\alpha$ exactly.
The former case means that we consider $H'$ as a perturbation from $H^\ell$ 
\begin{eqnarray}
H' = H^\ell + O(\sin \theta_{13}). \label{method-A}
\end{eqnarray}
We need to perform the first order perturbative calculation
to obtain $A$ and $B$ in this perturbative method.
Similarly, the later case means that we consider $H'$ 
as a perturbation from $H^h$
\begin{eqnarray}
H' = H^h + O(\alpha).  \label{method-B}
\end{eqnarray}
In order to calculate $A$ and $B$, we need to perform the first order 
perturbative calculation.
In both cases, we need to perform the first order perturbative 
calculation, because the CP violating effects disappear 
in the limit of vanishing $\alpha$ or $\theta_{13}$.  
\par
Let us interpret the above usual perturbative method
by using the general formulation (\ref{Coefficient-A}) and
(\ref{Coefficient-B}) as follows.
The expressions $A$ and $B$ are represented by two kinds of 
amplitudes $S'_{\mu e}$ and $S'_{\tau e}$. 
The dependence (\ref{order-S^h}) and (\ref{order-S^l}) 
of the two amplitudes on $\alpha$ and $\sin\theta_{13}$ 
is rewritten as 
\begin{eqnarray}
S'_{\mu e} = O(\alpha\sin^0 \theta_{13}),
 \quad S'_{\tau e} = O(\alpha^0\sin \theta_{13}). 
\end{eqnarray}
If we use the expansion in terms of $\sin\theta_{13}$ 
to calculate both amplitudes $S'_{\mu e}$ and $S'_{\tau e}$, 
the amplitude $S'_{\mu e}$ can be calculated in the zeroth 
order perturbation. 
However, $S'_{\tau e}$ need to be calculated in the first order 
perturbation.
In the same way, if we use the expansion in terms of $\alpha$,
the amplitude $S'_{\tau e}$ can be calculated in the zeroth 
order perturbation, but $S'_{\mu e}$ need to be calculated 
in the first order perturbation.
\par
An advantage of our method is that we are able to calculate 
both the amplitudes $S'_{\mu e}$ and $S'_{\tau e}$ 
{\it in the zeroth order perturbation}, namely  
{\it without the first order perturbation} of the small parameter 
$\alpha$ or $\sin\theta_{13}$.
If we expand $S'_{\mu e}$ in terms of 
$\sin\theta_{13}$ instead of $\alpha$, 
we do not need to perform the first order perturbation.
Similarly, if we expand $S'_{\tau e}$ in terms of 
$\alpha$ instead of $\sin\theta_{13}$, 
we do not need to perform the first order perturbation 
to calculate $S'_{\tau e}$.
One of the essential points of our method is that 
the Hamiltonian to calculate $S^\ell_{\mu e}$ 
is different from that to calculate $S^h_{\tau e}$.
Another point is that we only have to calculate $S^\ell_{\mu e}$ and 
$S^h_{\tau e}$ by using the Hamiltonian $H^\ell$ and $H^h$ 
in the framework of two generations, respectively.
These ideas make the calculations of the probability easy.

\section{Approximate Formula in Vacuum or in Constant Matter}

\hspace*{\parindent}
In this section, we calculate the concrete expressions for
$A, B$ and $C$ both in vacuum and in constant matter
by using the new method.
Moreover, we compare the value of these coefficients 
with exact value by numerical calculation.

\subsection{In Vacuum}

\hspace*{\parindent}
At first, we calculate $S^\ell_{\mu e}$ in vacuum, 
namely in the case of $a(t)=0$.
$S^\ell_{\mu e}$ is
\begin{eqnarray}
S^\ell_{\mu e} 
&=& \left(\exp(-iH^\ell L)\right)_{\mu e} \\
&=& \left(
O_{12}{\rm diag}(1, e^{-i\Delta_{21}L}, e^{-i\Delta_{31}L})O_{12}^T
\right)_{\mu e} \\
&=& -i \sin 2\theta_{12} \sin \frac{\Delta_{21}L}{2}
 \exp\left(-i\frac{\Delta_{21}}{2}L\right) \label{vacuum-S^l}
\end{eqnarray}
from the Hamiltonian (\ref{low-Hamiltonian}).
Similarly $S^h_{\tau e}$ is
\begin{eqnarray}
S^h_{\tau e} = -i \sin 2\theta_{13} \sin \frac{\Delta_{31}L}{2}
 \exp\left(-i\frac{\Delta_{31}}{2}L\right) \label{vacuum-S^h}
\end{eqnarray}
from the Hamiltonian (\ref{high-Hamiltonian}).
We obtain the expressions for $A, B$ and $C$ 
\begin{eqnarray}
A &\simeq& \sin 2\theta_{12}\sin 2\theta_{23}\sin 2\theta_{13}
\sin \frac{\Delta_{21}L}{2}
\sin \frac{\Delta_{31}L}{2}
\cos \frac{\Delta_{32}L}{2}, \\
B &\simeq& \sin 2\theta_{12}\sin 2\theta_{23}\sin 2\theta_{13}
\sin \frac{\Delta_{21}L}{2}
\sin \frac{\Delta_{31}L}{2}
\sin \frac{\Delta_{32}L}{2}, \\
C &\simeq& c_{23}^2\sin^2 2\theta_{12} 
\sin^2 \frac{\Delta_{21}L}{2}
+ s_{23}^2\sin^2 2\theta_{13} 
\sin^2 \frac{\Delta_{31}L}{2},
\end{eqnarray}
by substituting (\ref{vacuum-S^l}) and (\ref{vacuum-S^h}) 
into (\ref{approximate-A}), (\ref{approximate-B}) and
(\ref{approximate-C}).

\subsection{In Constant Matter}

\hspace*{\parindent}
Next we calculate the amplitudes in constant matter,
namely in the case of  $a(t)=a$. 
At first, we diagonalize the Hamiltonian (\ref{low-Hamiltonian})
in constant matter by the orthogonal matrix $O_{12}^\ell$
as 
\begin{eqnarray}
 H^\ell &=& O_{12}{\rm diag}(0,\Delta_{21},\Delta_{31}) O_{12}^T 
+ {\rm diag}(a,0,0) \\
&=& O^\ell_{12}{\rm diag}(\lambda_1^\ell,\lambda_2^\ell,\Delta_{31})
 (O_{12}^\ell)^T
\end{eqnarray}
to calculate $S_{\mu e}^\ell$. 
Here $\lambda_i^\ell (i=1,2)$ is the eigenvalue given by
\begin{eqnarray}
\lambda_i^\ell
= \frac{1}{2}
\left(
\Delta_{21} +a 
\pm \sqrt{(\Delta_{21}\cos 2\theta_{12}-a)^2 
+ \Delta_{21}^2\sin^2 2\theta_{12}}
\right), \label{eigenvalue-low}
\end{eqnarray}
and $\lambda_1^\ell$ and $\lambda_2^\ell$ correspond to the sign $-$ and 
the opposite sign $+$, respectively.
The effective mixing angle $\sin 2\theta^\ell_{12}$ is calculated as
\begin{eqnarray}
\sin^2 2\theta^\ell_{12} 
= \frac{\Delta_{21}^2\sin^2 2\theta_{12}}
{(\Delta_{21}\cos 2\theta_{12}-a)^2 
+ \Delta_{21}^2\sin^2 2\theta_{12}}. \label{eignvector-low}
\end{eqnarray}
From (\ref{eigenvalue-low}) and (\ref{eignvector-low}) we obtain the relation
\begin{eqnarray}
\frac{\Delta^\ell_{21}}{\Delta_{21}} 
= \frac{\sin 2\theta_{12}}{\sin 2\theta^\ell_{12}}
= \sqrt{\left(
      \cos 2\theta_{12}-\frac{a}{\Delta_{21}}
\right)^2 + \sin^2 2\theta_{12}}.
\end{eqnarray}
The amplitude $S^\ell_{\mu e}$ is calculated by using the $\lambda_i^\ell$ and
$\sin2\theta_{12}^\ell$ as
\begin{eqnarray}
S^\ell_{\mu e} 
&=& \left(\exp(-iH^\ell L)\right)_{\mu e} \\
&=& \left(
O_{12}{\rm diag}
(e^{-i\lambda_1^\ell L}, e^{-i\lambda_2^\ell L}, e^{-i\Delta_{31}L})
O_{12}^T
\right)_{\mu e} \\
&=& -i \sin 2\theta^\ell_{12} \sin \frac{\Delta^\ell_{21}L}{2}
 \exp\left(-i\frac{\lambda_1^\ell+\lambda_2^\ell}{2}L\right) \\
&=& -i \sin 2\theta^\ell_{12} \sin \frac{\Delta^\ell_{21}L}{2}
 \exp\left(-i\frac{\Delta_{21} +a}{2}L\right).
\label{constant-S^l}
\end{eqnarray}
Next let us calculate $S^h_{\tau e}$ from the Hamiltonian 
(\ref{high-Hamiltonian}) diagonalized by the orthogonal matrix 
$O_{13}^h$ 
\begin{eqnarray}
 H^h &=& O_{13}{\rm diag}(0,0,\Delta_{31})O_{13}^T 
+ {\rm diag}(a,0,0) \\
&=& O^h_{13}{\rm diag}(\lambda_1^h,0,\lambda_3^h)
 (O^h_{13})^T.
\end{eqnarray}
The eigenvalue $\lambda_i^h (i=1,3)$ of this Hamiltonian is given by
\begin{eqnarray}
\lambda_i^h
= \frac{1}{2}
\left(
\Delta_{31} +a 
\pm \sqrt{(\Delta_{31}\cos 2\theta_{13}-a)^2 
+ \Delta_{31}^2\sin^2 2\theta_{13}}
\right), \label{eigenvalue-high}
\end{eqnarray}
where $\lambda_1^h$ and $\lambda_3^h$ correspond to the sign $-$ and 
the opposite sign $+$.
Moreover we obtain the effective mixing angle 
$\sin 2\theta^h_{13}$ is calculated as
\begin{eqnarray}
\sin^2 2\theta^h_{13} 
= \frac{\Delta_{31}^2\sin^2 2\theta_{13}}
{(\Delta_{31}\cos 2\theta_{13}-a)^2 
+ \Delta_{31}^2\sin^2 2\theta_{13}}. \label{eignvector-high}
\end{eqnarray}
From (\ref{eigenvalue-high}) and (\ref{eignvector-high}) we obtain the relation
\begin{eqnarray}
\frac{\Delta^h_{31}}{\Delta_{31}} 
= \frac{\sin 2\theta_{13}}{\sin 2\theta^h_{13}}
= \sqrt{\left(
      \cos 2\theta_{13}-\frac{a}{\Delta_{31}}
\right)^2 + \sin^2 2\theta_{13}}.
\end{eqnarray}
The amplitude $S^h_{\tau e}$ is calculated as 
\begin{eqnarray}
S^h_{\tau e} = -i \sin 2\theta^h_{13} \sin \frac{\Delta^h_{31}L}{2}
 \exp \left(-i\frac{\Delta_{31} +a}{2}L \right). \label{constant-S^h}
\end{eqnarray}
Substituting (\ref{constant-S^l}) and (\ref{constant-S^h}) 
into (\ref{approximate-A}), (\ref{approximate-B}) and
(\ref{approximate-C}), we obtain
\begin{eqnarray}
A &\simeq& \sin 2\theta^\ell_{12}\sin 2\theta_{23}\sin 2\theta^h_{13}
\sin \frac{\Delta^\ell_{21}L}{2}
\sin \frac{\Delta^h_{31}L}{2}
\cos \frac{\Delta_{32}L}{2},
\label{constant-A} \\
B &\simeq& \sin 2\theta^\ell_{12}\sin 2\theta_{23}\sin 2\theta^h_{13}
\sin \frac{\Delta^\ell_{21}L}{2}
\sin \frac{\Delta^h_{31}L}{2}
\sin \frac{\Delta_{32}L}{2},
\label{constant-B} \\
C &\simeq& 
 c_{23}^2\sin^2 2\theta^\ell_{12} 
\sin^2 \frac{\Delta^\ell_{21}L}{2}
+ s_{23}^2\sin^2 2\theta^h_{13} 
\sin^2 \frac{\Delta^h_{31}L}{2}.
\label{constant-C}
\end{eqnarray}
The low and high energy MSW effects are contained 
in $\sin 2\theta^\ell_{12}, \Delta_{21}^\ell$ and  
in $\sin 2\theta^h_{13}, \Delta_{31}^h$ of the approximate formula,
respectively.
This is the reason why this approximate formula is applicable to 
a wide energy region.
The term including $B$, which is proportional to $\sin\delta$, 
is related to T violation  
\cite{Krastev-Petcov,Harrison-Scott,Yokomakura0009,Parke-Weiler}.
However it is difficult to observe only this term
in future long baseline experiments. 
Therefore there are many attempts to extract the information on 
the CP phase from the terms including both the coefficients $A$ and $B$ 
\cite{Lipari,Pinney,Minakata0108}.
\par
Next let us compare our approximate formula with the exact one.
We use the parameters $\Delta m_{21}^2 = 7.0 \times 10^{-5}$ eV$^2$, 
$\Delta m_{31}^2 = 2.0 \times 10^{-3}$ eV$^2$, 
$\sin^2 2\theta_{12}=0.8 $, $\sin^2 2\theta_{23}=1$,
$\sin \theta_{13}=0.16$, the oscillation length is $L = 730$ km
$a=\sqrt{2}G_FN_e$, where $G_F$ is the Fermi constant and 
$N_e$ is the electron density in matter calculated from 
the matter density $\rho=3 {\rm g/cm^3}$ and the electron fraction 
$Y_e=0.5$.
We plot the coefficients $A, B$ and $C$ as a function of the energy 
within the region $0.01 {\rm GeV} \leq E \leq 1 {\rm GeV}$.
These coefficients calculated from the exact formula 
are compared with those from the approximate formula in Fig. 1. 
\begin{figure}[ht]
  \begin{tabular}{ccc}
    $1(a)$ Exact & \quad $1(b)$ Ours & $1(c)$ Difference \\
    \resizebox{50mm}{!}{\includegraphics{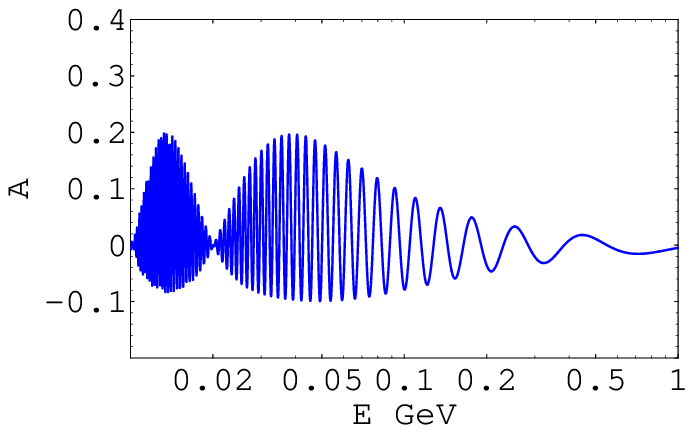}} &
    \resizebox{50mm}{!}{\includegraphics{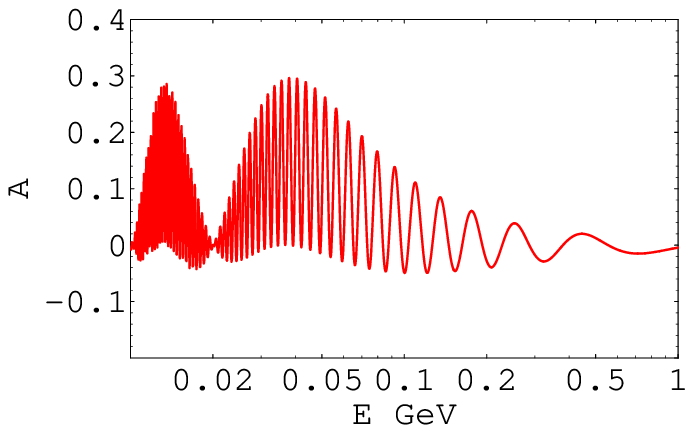}} &
    \resizebox{50mm}{!}{\includegraphics{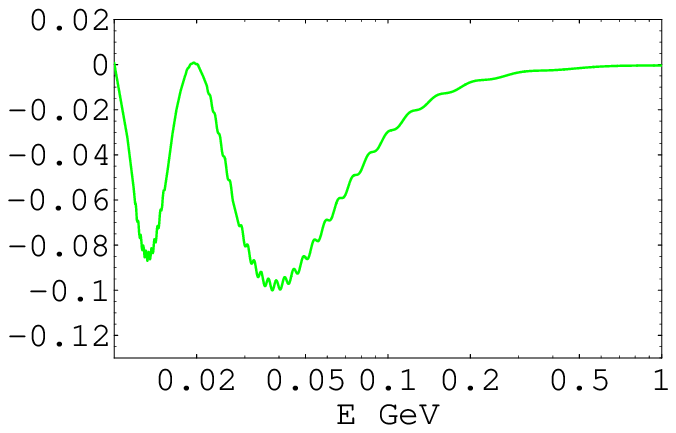}} \\
    $2(a)$ Exact & \quad $2(b)$ Ours & $2(c)$ Difference \\
    \resizebox{50mm}{!}{\includegraphics{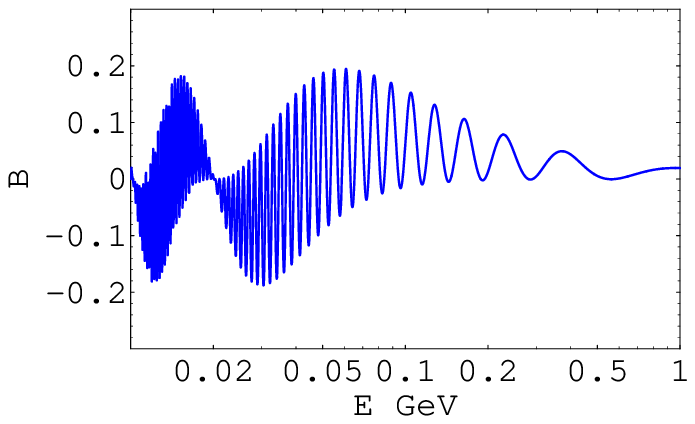}} &
    \resizebox{50mm}{!}{\includegraphics{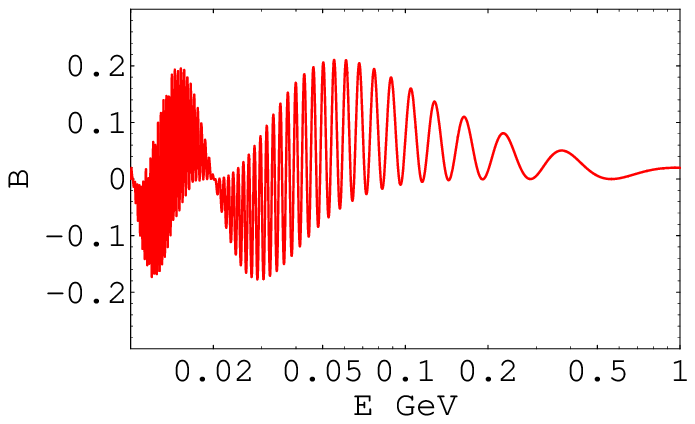}} &
    \resizebox{50mm}{!}{\includegraphics{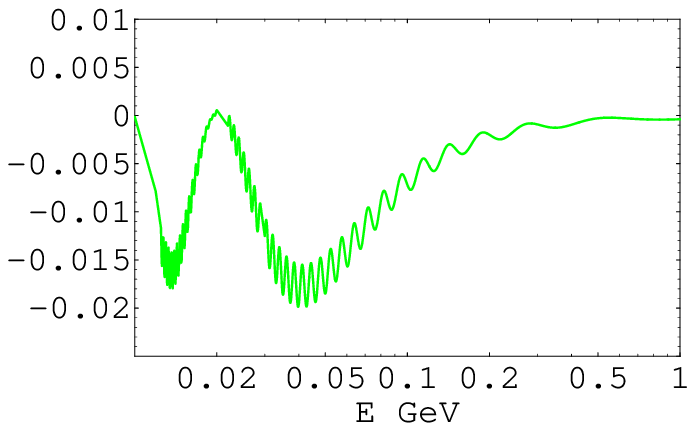}} \\
    $3(a)$ Exact & \quad $3(b)$ Ours & $3(c)$ Difference \\
    \resizebox{50mm}{!}{\includegraphics{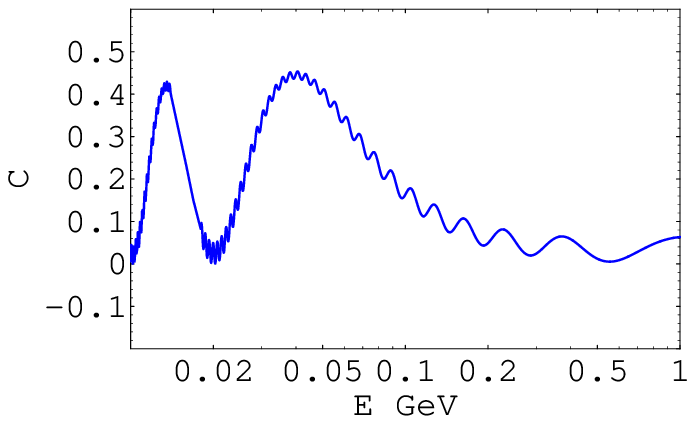}} &
    \resizebox{50mm}{!}{\includegraphics{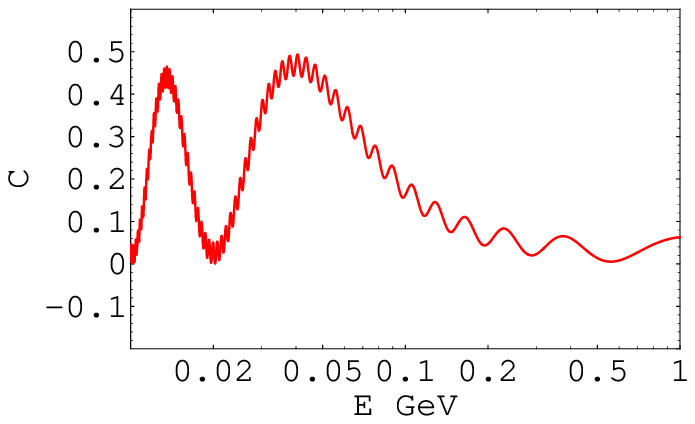}} &
    \resizebox{50mm}{!}{\includegraphics{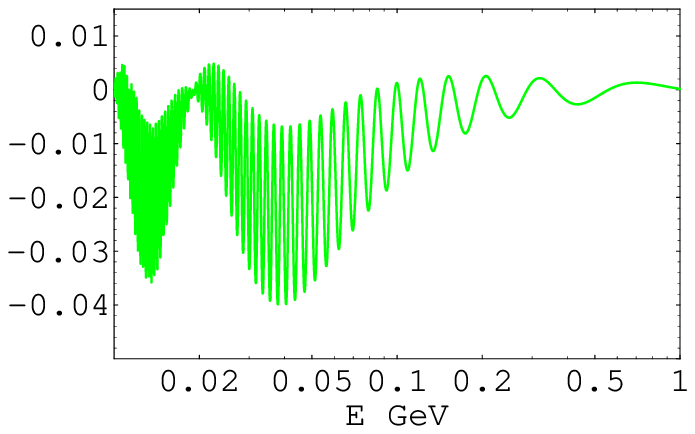}} \\
  \end{tabular}
\caption[l]{Comparison of our reduction formula with the exact one
in the coefficients $A,B$ and $C$, plotted from top to bottom and 
the exact, approximate formula and the difference from left to right.}
\end{figure}
From this figure we find that the approximate formula almost coincide 
with the exact formula. 
The error is estimated to be less than 20 $\%$ from Fig. 1. 
The difference between the exact and the approximate formulas 
is caused by ignoring the higher order terms 
in the perturbative expansion on $\sin \theta_{13}$ 
and $\alpha$.

\section{Derivation of Other Approximate Formulas}

\hspace*{\parindent}
In this section, we derive the approximate formulas 
given by other authors in constant matter from our formula.
There are formulas for the low energy 
\cite{Koike2000, Yasuda1999, Minakata2000},
the intermediate energy \cite{AKS,AS,Sato2000} 
and the high energy regions \cite{Freund2000, Cervera, Freund}.

\subsection{Low Energy Formula}

\hspace*{\parindent}
At first we derive a low energy formula with large mixing angle 
$\theta_{12}$, which is similar to those 
in \cite{Koike2000, Yasuda1999, Minakata2000}.
Under the low energy condition 
\begin{eqnarray}
a \ll \Delta_{31}, \label{lowenergy}
\end{eqnarray} 
the following relation
\begin{eqnarray}
\frac{\Delta^h_{31}}{\Delta_{31}} 
= \frac{\sin 2\theta_{13}}{\sin 2\theta^h_{13}}
\simeq 1 \label{ratio31}
\end{eqnarray}
is derived by expanding $\Delta^h_{31}$ and  $\sin 2\theta^h_{13}$ 
in terms of $a/\Delta_{31}$. 
Namely $\Delta^h_{31}$ and  $\sin 2\theta^h_{13}$ in matter
can be approximated by the quantities in vacuum.
Furthermore, if we take the limit $\theta_{12} \to \pi/4$ 
\begin{eqnarray}
\frac{\Delta^\ell_{21}}{\Delta_{21}} 
= \frac{\sin 2\theta_{12}}{\sin 2\theta^\ell_{12}}
\simeq \frac{\sqrt{\Delta_{21}^2+a^2}}{\Delta_{21}}
\label{ratio21}
\end{eqnarray}
is obtained. By using the relations (\ref{ratio31}) and
(\ref{ratio21}), the coefficients $A,B$ and $C$ for 
(\ref{constant-A}), (\ref{constant-B}) and (\ref{constant-C})
are reduced to the following expressions
\begin{eqnarray}
A &\simeq& \frac{\Delta_{21}}{\sqrt{\Delta_{21}^2+a^2}}
\sin 2\theta_{12}\sin 2\theta_{23}\sin 2\theta_{13}
 \sin \frac{\Delta_{31}L}{2}\sin 
 \frac{\sqrt{\Delta_{21}^2+a^2}L}{2}
 \cos \frac{\Delta_{32}L}{2}, \label{low-A} \\ 
B &\simeq& \frac{\Delta_{21}}{\sqrt{\Delta_{21}^2+a^2}}
\sin 2\theta_{12}\sin 2\theta_{23}\sin 2\theta_{13}
 \sin \frac{\Delta_{31}L}{2}\sin 
 \frac{\sqrt{\Delta_{21}^2+a^2}L}{2}
 \sin \frac{\Delta_{32}L}{2}, \label{low-B} \\ 
C &\simeq& 
\frac{\Delta_{21}^2}{\Delta_{21}^2+a^2}c_{23}^2\sin^2 2\theta_{12}
\sin^2  \frac{\sqrt{\Delta_{21}^2+a^2}L}{2}
+ s_{23}^2 \sin^2 2\theta_{13}\sin^2 \frac{\Delta_{31}L}{2},
\label{low-C} 
\end{eqnarray}
where the condition derived 
from the low energy condition (\ref{lowenergy})
\begin{eqnarray}
\sin \frac{\Delta_{31}^hL}{2} 
\simeq \sin \frac{\Delta_{31}L}{2} 
\label{shortbaseline-approximation-3}
\end{eqnarray}
is also used. The applicable region for energy is given by
\begin{eqnarray}
E \ll 15 \: {\rm GeV} 
\left(
  \frac{\Delta m_{31}^2}{10^{-3}\:{\rm eV^2}}
\right)
\left(
  \frac{3\:{\rm g/cm^3}}{\rho}
\right) \label{lowenergy-2}
\end{eqnarray}
from the condition (\ref{lowenergy}). 
In addition to this condition, the applicable region of 
(\ref{low-A}), (\ref{low-B}) and (\ref{low-C}) is 
restricted by
\begin{eqnarray}
 L \ll 8000 \: {\rm km} 
\left(
  \frac{E}{{\rm GeV}}
\right)
\left(
  \frac{10^{-4}\:{\rm eV^2}}{\Delta m_{21}^2}
\right),
\end{eqnarray}
which comes from the approximation in the oscillation parts
of (\ref{low-A}), (\ref{low-B}) and (\ref{low-C}).
A similar result can be obtained for the perturbation of 
$\sin \theta_{13}$ \cite{Akhmedov}.
They have proposed a low energy formula in arbitrary matter,
by using the first order perturvative calculations. 
Our method has the advantage that the calculation is 
much simpler.
This approximate formula coincides with that in vacuum 
in the low energy limit, or in other words, 
this result recovers the vacuum mimicking phenomenon
which has been discussed in \cite{Minakata2000,Yasuda0106}.

\subsection{Intermediate Energy Formula}

\hspace*{\parindent}
At first we derive the intermediate energy formula \cite{AKS,AS,Sato2000}
from our formula.
Under the low energy condition
\begin{eqnarray}
a \ll \Delta_{31},
\end{eqnarray}
we expand $\Delta^h_{31}$ and $\sin 2\theta^h_{13}$ 
up to first order of $a/\Delta_{31}$
\begin{eqnarray}
\Delta^h_{31} 
&\simeq& \Delta_{31} - 2a \cos 2\theta_{13}, \label{mass31} \\
\sin 2\theta^h_{13} 
&\simeq& \sin 2\theta_{13}
\left(1 + \frac{2a}{\Delta_{31}}\cos 2\theta_{13}\right).\label{mixing13}
\end{eqnarray}
Substituting (\ref{mass31}) and (\ref{mixing13}) into
(\ref{constant-S^h}), we obtain the expression
\begin{eqnarray}
|S^h_{\tau e}|^2 
&\simeq& s_{23}^2\sin^2 2\theta_{13}
\left(1 + \frac{2a}{\Delta_{31}}\cos 2\theta_{13}\right)
\sin^2 \frac{(\Delta_{31} - 2a \cos 2\theta_{13})L}{2} \\
&\simeq& s_{23}^2\sin^2 2\theta_{13}
\left[
\left(1 + \frac{2a}{\Delta_{31}}\cos 2\theta_{13}\right)
\sin^2 \frac{\Delta_{31}L}{2}
- aL\cos 2\theta_{13} \sin (\Delta_{31}L)
\right],
\end{eqnarray}
where we also use the approximation 
\begin{eqnarray}
aL\cos 2\theta_{13} \ll 1 \label{shorbaseline-approximation}
\end{eqnarray}
from the first line to the second line.
Furthermore, under the assumption that $\Delta_{21}L/2$ is small,
we can approximate   
\begin{eqnarray}
\sin \frac{\Delta_{21}^\ell L}{2}
\simeq \frac{\Delta_{21}^\ell L}{2}
\simeq \frac{\Delta_{21}L}{2}, \label{common-approximation}
\end{eqnarray}
and from (\ref{constant-A}), (\ref{constant-B}),
(\ref{constant-C}), the approximate formula for $A, B$ and $C$ is
derived as 
\begin{eqnarray}
A &\simeq& 
\frac{1}{2}\sin 2\theta_{12}\sin 2\theta_{23}\sin 2\theta_{13}
\frac{\Delta_{21}L}{2}\sin (\Delta_{31}L), \\
B &\simeq& 
\sin 2\theta_{12}\sin 2\theta_{23}\sin 2\theta_{13}
\frac{\Delta_{21}L}{2}\sin^2 \frac{\Delta_{31}L}{2}, \\
C &\simeq& 
s_{23}^2\sin^2 2\theta_{13}
\left[
\left(1 + \frac{2a}{\Delta_{31}}\cos 2\theta_{13}\right)
\sin^2 \frac{\Delta_{31}L}{2}
- aL\cos 2\theta_{13} \sin (\Delta_{31}L)
\right].
\end{eqnarray}
One of the conditions for the applicable region of this approximate 
formula, namely for the upper limit 
\begin{eqnarray}
E \ll 15 \: {\rm GeV} 
\left(
  \frac{\Delta m_{31}^2}{10^{-3}\:{\rm eV^2}}
\right)
\left(
  \frac{3\:{\rm g/cm^3}}{\rho}
\right) \label{lowenergy-3}
\end{eqnarray}
is the same as that in the low energy region (\ref{lowenergy}).
In addition to this, the conditions due to
(\ref{shorbaseline-approximation}) and 
(\ref{common-approximation})
\begin{eqnarray}
L &\ll& 1700 \: {\rm km} 
\left(
  \frac{3\:{\rm g/cm^3}}{\rho}
\right), \label{shorbaseline-approximation-2} \\
E &\gg& 0.185 \: {\rm GeV} 
\left(
  \frac{\Delta m_{21}^2}{10^{-4}\:{\rm eV^2}}
\right)
\left(
  \frac{L}{730\:{\rm km}}
\right) \label{common-approximation-2}
\end{eqnarray}
should be satisfied. 
This approximate formula has been derived by using the perturbations
of $\alpha$ and $a/\Delta m^2_{31}$ \cite{AKS}.
From (\ref{lowenergy-3}), (\ref{shorbaseline-approximation-2}) 
and (\ref{common-approximation-2}), the applicable region 
is rather restricted because of the expansion of the oscillation part.
On the other hand, it has the advantage that the contribution 
of the genuine CP violation can be easily distinguished from 
that of the fake CP violation.

\subsection{High Energy Formula}

\hspace*{\parindent}
Next, we derive the high energy formulas 
\cite{Freund2000,Cervera, Freund} from our formula.
Under the high energy condition 
\begin{eqnarray}
a \gg \Delta_{21}, \label{highenergy}
\end{eqnarray} 
we obtain
\begin{eqnarray}
\frac{\Delta^\ell_{21}}{\Delta_{21}} 
= \frac{\sin 2\theta_{12}}{\sin 2\theta^\ell_{12}}
\simeq \frac{a}{\Delta_{21}} \label{ratio21h}
\end{eqnarray}
by expanding $\Delta^h_{21}$ and $\sin 2\theta^h_{12}$ 
up to the first order of $\Delta_{21}/a$.
In addition, using the approximation $\theta_{13} \to 0$, we obtain
another relation 
\begin{eqnarray}
\frac{\Delta^h_{31}}{\Delta_{31}} 
= \frac{\sin 2\theta_{13}}{\sin 2\theta^h_{13}}
\simeq 1-\frac{a}{\Delta_{31}}. \label{ratio31h}
\end{eqnarray}
The concrete expressions of $A, B$ and $C$ are derived as
\begin{eqnarray}
A &\simeq& \frac{\Delta_{21}\Delta_{31}}{a(\Delta_{31}-a)}
\sin 2\theta_{12}\sin 2\theta_{23}\sin 2\theta_{13}
 \sin \frac{aL}{2}\sin \frac{(\Delta_{31}-a)L}{2}
 \cos \frac{\Delta_{32}L}{2}, \label{high-a} \\
B &\simeq& \frac{\Delta_{21}\Delta_{31}}{a(\Delta_{31}-a)}
\sin 2\theta_{12}\sin 2\theta_{23}\sin 2\theta_{13}
 \sin \frac{aL}{2}\sin \frac{(\Delta_{31}-a)L}{2}
 \sin \frac{\Delta_{32}L}{2}, \label{high-b} \\
C &\simeq& 
\frac{\Delta_{21}^2}{a^2}c_{23}^2\sin^2 2\theta_{12} 
\sin^2 \frac{aL}{2} 
+ \frac{\Delta_{31}^2}{(\Delta_{31}-a)^2}s_{23}^2\sin^2 2\theta_{13}
\sin^2 \frac{(\Delta_{31}-a)L}{2},\label{high-c}
\end{eqnarray}
by substituting (\ref{ratio21h}) and (\ref{ratio31h})
into (\ref{constant-A}), (\ref{constant-B}) and
(\ref{constant-C}), where we also use the approximation
\begin{eqnarray}
\sin \frac{\Delta_{21}^\ell L}{2} \simeq \sin \frac{aL}{2}. 
\label{shortbaseline-approximation-4}
\end{eqnarray}
The applicable region of this approximate formula 
is calculated from (\ref{highenergy}) as
\begin{eqnarray}
E \gg 0.45 \: {\rm GeV} 
\left(
  \frac{\Delta m_{21}^2}{10^{-4}\:{\rm eV^2}}
\right)
\left(
  \frac{3\:{\rm g/cm^3}}{\rho}
\right).
\end{eqnarray}
In addition to this, the applicable region is also restricted by
\begin{eqnarray}
 L \ll 8000 \: {\rm km} 
\left(
  \frac{E}{{\rm GeV}}
\right)
\left(
  \frac{10^{-4}\:{\rm eV^2}}{\Delta m_{21}^2}
\right),
\end{eqnarray}
which is derived from (\ref{shortbaseline-approximation-4}). 
Although these high energy approximate formulas 
(\ref{high-a}), (\ref{high-b}) and (\ref{high-c})
have been derived at first in \cite{Cervera, Freund},
their derivation is complicated because of the
calculation up to the first order perturbation of
$\alpha$.
Here, we have presented the simple derivation of these formulas
by using the new idea of taking only the zeroth order perturbation.

\section{Summary}

\hspace*{\parindent}
In this paper, we study the oscillation probability in matter 
within the framework of three generations.
The results are as follows.
\begin{enumerate}
\item[1.]We have proposed a simple method to approximate 
the oscillation probability in arbitrary matter. 
Our method provide an approximate formula in arbitrary matter 
{\it without the usual first order perturbative calculations}
of the small parameter $\Delta m_{21}^2/\Delta m_{31}^2$
or $\sin \theta_{13}$.
\item[2.]
The concrete expressions for our approximate formula 
in constant matter has been derived to investigate 
the accuracy of the reduction formula 
(\ref{approximate-P})-(\ref{approximate-C}).
We have shown that our formula is numerically in good agreement 
with the exact solution with reasonable accuracy.
\item[3.]
We have shown that both the low energy 
\cite{Koike2000, Yasuda1999, Minakata2000},
the intermediate energy \cite{AKS,AS,Sato2000}
and the high energy \cite{Freund2000,Cervera, Freund} 
approximate formulas in constant matter presented by other authors 
can be easily derived from our formula.
This means that our formula is applicable to a wide 
energy region.
\end{enumerate}

\vspace{20pt}
\noindent
{\Large {\bf Acknowledgment}}

\noindent
We would like to thank Prof. Wilfried Wunderlich for helpful
comments and advice on English expressions.

\end{document}